\documentstyle[12pt,epsf]{article}
\newdimen\rotdimen
\def\vspec#1{\special{ps:#1}}
\def\rotstart#1{\vspec{gsave currentpoint currentpoint translate
   #1 neg exch neg exch translate}}
\def\rotfinish{\vspec{currentpoint grestore moveto}}
\def\rotr#1{\rotdimen=\ht#1\advance\rotdimen by\dp#1%
   \hbox to\rotdimen{\hskip\ht#1\vbox to\wd#1{\rotstart{90 rotate}%
   \box#1\vss}\hss}\rotfinish}
\def\rotl#1{\rotdimen=\ht#1\advance\rotdimen by\dp#1%
   \hbox to\rotdimen{\vbox to\wd#1{\vskip\wd#1\rotstart{270 rotate}%
   \box#1\vss}\hss}\rotfinish}%
\def\rotu#1{\rotdimen=\ht#1\advance\rotdimen by\dp#1%
   \hbox to\wd#1{\hskip\wd#1\vbox to\rotdimen{\vskip\rotdimen
   \rotstart{-1 dup scale}\box#1\vss}\hss}\rotfinish}%
\def\rotf#1{\hbox to\wd#1{\hskip\wd#1\rotstart{-1 1 scale}%
   \box#1\hss}\rotfinish}%

\begin{document}

\author{A. J. Buchmann and Amand Faessler \\
{\it Instit\"ut f\"ur Theoretische Physik, Universit\"at T\"ubingen}\\
{\it Auf der Morgenstelle 14, D-72076 T\"ubingen, Germany} \and M. I.
Krivoruchenko \\
{\it Institute for Theoretical and Experimental Physics }\\
{\it \ B.Cheremushkinskaya 25, 117259 Moscow, Russia}}
\title{DIBARYON CONDENSATE IN NUCLEAR MATTER AND NEUTRON STARS: EXACT ANALYSIS IN
ONE-DIMENSIONAL MODELS}
\date{}
\maketitle

\begin{abstract}
Phase transitions of nuclear matter to the quark-gluon plasma with
subsequent restoration of chiral symmetry have been widely discussed in the
literature. We investigate the possibility for occurrence of dense nuclear
matter with a dibaryon Bose-Einstein condensate as an intermediate state
below the quark-gluon phase transition. An exact analysis of this state of
matter is presented in a one-dimensional model. The analysis is based on
a reduction of the quantization rules for the $N$-body problem to $N$
coupled algebraic transcendental equations. We observe that when the Fermi
momentum approaches the resonance momentum, the one-particle distribution
function increases near the Fermi surface. When the Fermi momentum is
increased beyond the resonance momentum, the equation of state becomes
softer. The observed behavior can be interpreted in terms of formation of a
Bose-Einstein condensate of two-fermion resonances (e.g. dibaryons). In cold
nuclear matter, it should occur at $2(m_N+\varepsilon _F)\geq m_D$ where $
m_N $ and $m_D$ are the nucleon and dibaryon masses and $\varepsilon _F$ is
the nucleon Fermi energy.
\end{abstract}

\section{INTRODUCTION}

\setcounter{equation}{0}

Exactly solvable models are of considerable interest. They provide important
tests of different approximation schemes. These approximation methods may
then be applied with increased confidence to more complex cases where
analytic solutions are impossible. In fact, one-dimensional models very
often provide the only possibility to gain analytic insights into the
behavior of an interacting many-body system that go beyond perturbation
theory. In this sense, exactly solvable low-dimensional models may provide
valuable guidance in finding the proper dynamical description of more
complicated systems.

The recent discovery \cite{And} of Bose-Einstein condensation in a dilute
vapor of rubidium-87 atoms created new interest in the phenomenon of
Bose-Einstein condensation and in exactly solvable models of interacting
bosons. Already some time ago \cite{Gir} it was noted that the energy
spectrum of a one-dimensional Bose-gas of nonpenetrable particles is
identical to that of an ideal Fermi gas. In other words, there is an exact
one-to-one correspondence between bosons which interact via an ''infinite
wall'' two-body potential, and a system of noninteracting fermions. Later 
\cite{Lie1,Lie2} an exact analytic solution for a one-dimensional Bose-gas
interacting via a delta-function potential $V(x)=\alpha \delta (x)$, where $
\alpha $ is an arbitrary positive parameter was obtained. It was shown that
the model of ref. [2] is a special case of the model in refs. \cite
{Lie1,Lie2}, since in the limit $\alpha \rightarrow \infty $\ particles do
not penetrate each other. For $\alpha <\infty $ the qualitative features of
the energy spectrum remain unaltered. A class of exact solutions 
three particles with different masses interacting through a
finite-strength delta-function potential was found \cite{McG}. Thermodynamic
properties of an interacting Bose-gas with the potential $V(x)=\alpha \delta
(x)$ were also discussed \cite{Yan}. It was demonstrated that at finite
temperatures the thermodynamic functions do not show a non-analytic
behavior. In ref. \cite{Son} the possibility of superfluidity at zero
temperature is discussed. In ref. \cite{Bru} it was shown that long-range
correlations inherent in the models of refs. \cite{Gir,Lie1} decrease as
some power of $1/x$ at $T=0$. One-dimensional models have also led to a
deeper understanding of critical phenomena \cite{Fra}. Exact results for
the quantization of the Toda lattice are reviewed in ref. [10].

There is an extensive literature devoted to
dibaryon resonances. Dibaryons are predicted by most low-energy models 
of quantum chromodynamics (QCD) [11-14]. Some candidates are reported to be 
seen in experiments \cite{Par,Bil}. The appearance of Bose-Einstein 
condensation of
two-fermion resonances in nuclear matter softens the equation of state of
nuclear matter and decreases the upper limits for masses of neutron stars 
\cite{Kri}. In the framework of the Walecka model, heterogenous
nucleon-dibaryon matter is discussed in ref. \cite{Fae}. In the present
paper, we focus on one-dimensional Fermi-systems with a Bose-resonance in
the two-fermion channel. These one-dimensional models might have important
implications for the behavior of dibaryon resonances in the nuclear medium.

Previous results [2-8] were obtained for a $\delta $-function interaction
potential. Here, we show that one-dimensional models admit exact solutions
for other types of zero-range singular potentials. We classify
eigenfunctions of the $N$-body Hamiltonian, establish quantization rules,
find eigenvalues of the $N$-body Hamiltonian without recourse to
perturbation theory. In the thermodynamic limit one can exactly
(numerically) calculate dispersion laws for elementary excitations, the
ground state energy, and the equation of state.

The outline of the paper is as follows. We start with discussing the simpler
case of a system of bosons interacting through an arbitrary finite-range
potential $V(x)$ ($V(x)\neq 0$ for $\mid x\mid \leq a$ and $V(x)=0$ for $
\mid x\mid >a$) generating a nontrivial two-body scattering $S$-matrix.
We then pass to the limit 
$V(x)\rightarrow \infty $ for $\mid x\mid \leq a$ and 
$a\rightarrow 0$. In this limit, there is a wide class of nontrivial finite
two-body $S$-matrices. In sect. 2, we establish the properties of the
coefficients entering the plane wave expansion of the $N$-boson wave
function. In particular, we study their properties with respect to
permutations of the particle quasi-momenta and establish relations between
these coefficients. Then we discuss the symmetry of the wave function under
permutations of the arguments, periodic boundary conditions, and matching
conditions for the wave function. In Sect. 3, we derive the generalized
Bohr-Sommerfeld quantization rules which reduce the quantization problem to
the problem of finding solutions to $N$ coupled algebraic transcendental
equations. In the limit $a\rightarrow 0$, the energy eigenvalues are
completely determined by the two-particle scattering phase shifts. In Sect.
4, we discuss the thermodynamic limit $N$ $\rightarrow \infty $, $
L\rightarrow \infty $, $N/L=$ $constant$. A linear integral equation is
derived for the distribution function of the quasi-momenta of the
particles. The distribution function is completely determined by the
scattering phase shifts of the particles. The elementary excitations in the
Bose-system are classified and their dispersion laws are established. 
In Sect. 5, the results of Sects. 2 and 3 are
extended to special classes of exactly solvable Fermi-systems. In Sect. 6,
we discuss the properties of a system of fermions interacting via potentials
which allow for a resonance in the two-body $S$-matrix. In Sect. 7 the
problem of stability of self-gravitating objects (neutron stars) made up of
fermions with a resonance interaction is discussed. In one dimension all
objects of such a kind are stable in general. In three-dimensional space
narrow resonances cause instability of massive neutron stars. Finally, in
sect. 8 we summarize and discuss the results.

\-

\section{EXPANSION COEFFICIENTS OF THE WAVE FUNCTION}

\setcounter{equation}{0}

We search for solutions of the $N$-body Schr\"odinger equation in one
dimension 
\begin{equation}
\label{II.1}(\sum_j^N(\frac{\hat p_j^2}{2m}+\sum_{i<j}^{}V(x_i-x_j))\Psi
(x_1,...,x_N)=E\Psi (x_1,...,x_N) 
\end{equation}
in the interval $[0,L]$. Here, the particles with mass $m$ are assumed to be
bosons, so the wave function $\Psi (x_1,...,x_N)$ is symmetric under
permutation of any pair of its arguments. The potential $V(x_i-x_j)$ is
assumed to vanish for $|x_i-x_j|>a$. Finally, we pass to the limit $
V(x)\rightarrow \infty $ for $\mid x\mid <a$ and $a\rightarrow 0.$ In this
limit there exists a wide class of nontrivial two-body $S$-matrices.

The problem considered here is quite similar to the one-dimensional
Heisenberg model of ferromagnetism with nearest neighbor interactions, first
solved by Bethe \cite{Bet}.

Following refs. \cite{Gir,Lie1}, we impose periodic boundary conditions for
the wave function 
\begin{equation}
\label{II.2}\Psi (x_1,...,x_j=0,...,x_N)=\Psi (x_1,...,x_j=L,...,x_N) 
\end{equation}
for any $j$.

One can verify that the function 
\begin{equation}
\label{II.3}\chi (x_1,...,x_N)=\exp (i\sum_j^Nk_jx_j) 
\end{equation}
satisfies the Schr\"odinger equation, if $x_j+a<x_{j+1}$, i.e. when $
V(x_i-x_j)=0$ for any pair of the arguments. The energy equals 
\begin{equation}
\label{II.4}E=\sum_j^N\frac{k_j^2}{2m}. 
\end{equation}

It is evident that any function which differs from Eq. (2.3) by a
permutation of the particle quasi-momenta $(k_1,...,k_N)$ $\rightarrow $\ $
(k_{\alpha _1},...,k_{\alpha _N})$ satisfies the same equation and has the
same energy. The problem reduces therefore to (i) the determination of the
weights of all components $(k_{\alpha _1},...,k_{\alpha _N})$ of the wave
function $\Psi (x_1,...,x_N)$, (ii) matching the wave functions in the
different regions of integrability (''$A$-regions'') 
where the exact wave function can be represented as a 
superposition of plane waves (2.3), (iii) symmetrization of the
expression (2.3) with respect to the arguments, (iv) taking into account
periodic boundary conditions, and finally (v) derivation of the
multidimensional analog of the Bohr-Sommerfeld quantization rules for the
particle quasi-momenta $k_i$.

Let us order the arguments of the wave function in increasing sequence, e.g. 
$x_5<x_1<x_3<...<x_{12}$. The numbers $(5,1,3,...,12)$ constitute a set $
(\gamma _1,...,\gamma _N)$. Therefore, $\gamma _i$ is the number of the
argument that occupies the $i$-th place in the above ordered sequence. Each $
A$-region can be brought into correspondence with a set $(\gamma
_1,...,\gamma _N)$ which is a permutation of the numbers $(1,...,N)$. There
are $N!$ regions in which solutions can be represented in form of plane
waves. Each $A$-region is fixed by a set of inequalities 
\begin{equation}
\label{II.5}x_{\gamma _j}+a<x_{\gamma _{j+1}} 
\end{equation}
for $j=1,$ $N-1$, so that $V(x_i-x_j)=0$ for any pair of the arguments. The
solutions of the Schr\"odinger equation in the region $(\gamma _1,...,\gamma
_N)$ can be written in the form 
\begin{equation}
\label{II.6}\Psi (x_1,...,x_N)=\sum_{\alpha _1...\alpha _N}C_{\alpha
_1...\alpha _N}^{\gamma _1...\gamma _N}\exp (i\sum_j^Nk_jx_j). 
\end{equation}
The sum is taken over the $N!$ permutations of the $(k_1,...,k_N)$. It is
assumed that the $k$'s are all different. It will be shown below that this
is a general case.

The wave function $\Psi (x_1,...,x_N)$\ contains for finite-range potentials
terms with different sets of the particle quasi-momenta $\{k_i\}.$ In the
limit $a\rightarrow 0$, however, only one unique set $\{k_i\}$ survives. The
requirement $a\rightarrow 0$ is necessary to ensure completeness of the
plane wave expansion in the $A$-regions.

Now, we consider the periodic boundary conditions. Suppose the set of
arguments of the wave function on the left hand side of Eq. (2.2) belongs to
the region $(\gamma _1,...,\gamma _N)$, then $\gamma _1=j$ due to the
condition $x_j=0$. The set of arguments of the wave function on the right
hand side of the Eq. (2.2) belongs to the region $(\lambda _1,...,\lambda _N)
$.

It is clear that $\lambda _i=\gamma _i+1$ for $i<N$, $\lambda _N=\gamma _1$
by virtue of $x_j=L$. The terms of the sum (2.6) are all linearly
independent. Therefore, the periodic boundary condition (2.2) can be
unambiguously projected to the expansion coefficients 
\begin{equation}
\label{II.7}C_{\alpha _1...\alpha _{N-1},\alpha _N}^{\gamma _1...\gamma
_{N-1},\gamma _N}=C_{\alpha _2...\alpha _N,\alpha _1}^{\gamma _2...\gamma
_N,\gamma _1}\exp (ik_{\alpha _1}L). 
\end{equation}

We illustrate this with an example. Let $N=2$, $j=1$ then 
$$
\begin{array}{c}
\Psi (0,x_2)=C_{12}^{12}\exp (ik_2x_2)+C_{21}^{12}\exp (ik_1x_2), \\ 
\Psi (L,x_2)=C_{12}^{21}\exp (ik_1x_2+ik_2L)+C_{21}^{21}\exp
(ik_2x_2+ik_1L). 
\end{array}
$$
From the condition $\Psi (0,x_2)=\Psi (L,x_2)$ follows $
C_{12}^{12}=C_{21}^{21}\exp (ik_1L)$ and $C_{21}^{12}=C_{12}^{21}\exp
(ik_2L) $.

Under exchange of the coordinates $x_{\gamma _s}\leftrightarrow
x_{\gamma r}$ one region of integrability transforms to another one. The
symmetry conditions for the wave functions under these permutations can be
formulated in terms of the expansion coefficients in the following way 
\begin{equation}
\label{II.8}C_{\alpha _1...\alpha _s...\alpha _r...\alpha _N}^{\gamma
_1...\gamma _s...\gamma _r...\gamma _N}=C_{\alpha _1...\alpha _s...\alpha
_r...\alpha _N}^{\gamma _1...\gamma _r...\gamma _s...\gamma _N}. 
\end{equation}

Let us illustrate this again with an example. For $N=2$, $x_1<x_2$ we get 
$$
\begin{array}{c}
\Psi (x_1,x_2)=C_{12}^{12}\exp (ik_1x_1+ik_2x_2)+C_{21}^{12}\exp
(ik_2x_1+ik_1x_2), \\ 
\Psi (x_2,x_1)=C_{12}^{21}\exp (ik_1x_1+ik_2x_2)+C_{21}^{21}\exp
(ik_2x_1+ik_1x_2). 
\end{array}
$$
The condition $\Psi (x_1,x_2)=\Psi (x_2,x_1)$ implies $
C_{12}^{12}=C_{12}^{21},$ $C_{21}^{12}=C_{21}^{21}$.

We consider the matching conditions for the wave functions in different
A-regions. Consider first the two regions $\Gamma _{\pm }$ which can be
obtained from each other by a permutation of the arguments $x_{\gamma _s}$
and $x_{\gamma r}$ such that $s\pm 1=r$. In the sequence of increasing
arguments of the wave function, the values $x_{\gamma _s}$ and $x_{\gamma r}$
occupy neighboring places. In the region $\Gamma _{+}$, $x_{\gamma _s}$ is
on the right, i.e. $s+1=r$, while in the region $\Gamma _{-}$, $x_{\gamma r}$
is on the left, i.e. $s-1=r$. Let $\xi _1=x_{\gamma s}$, $\xi _2=x_{\gamma r}
$. We now join the regions $\Gamma _{+}$ and $\Gamma _{-}$ and remove the
restriction $|\xi _1-\xi _2|>a$. In the region obtained (denoted herewith by 
$\Gamma $), we are looking for solutions of the Schr\"odinger equation in
the form 
\begin{equation}
\label{II.9}\exp (i\sum_j^Nk_jx_j)\chi (\xi _1,\xi _2).
\end{equation}
The sum is extended over $j\neq s,r$. Let $q_1=k_{\alpha s}$, $q_2=k_{\alpha
r}$. In the region $\Gamma $, the function $\chi (\xi _1,\xi _2)$ satisfies
the equation 
\begin{equation}
\label{II.10}(\frac{\hat p_1^2}{2m}+\frac{\hat p_2^2}{2m}+V(\xi _1-\xi
_2))\chi (\xi _1,\xi _2)=E^{\prime }\chi (\xi _1,\xi _2)
\end{equation}
with $E^{\prime }=\frac{q_1^2}{2m}+\frac{q_2^2}{2m}$. The interval $|\xi
_1-\xi _2|<a$, in which the potential is different from zero, is interesting
only for supplying the correct matching conditions for the wave function for 
$\xi _2-\xi _1>a$ and $\xi _2-\xi _1<-a$. In terms of the total and relative
quasi-momenta of the particles, $K=q_1+q_2$, $k=(q_2-q_1)/2$, the
center-of-mass coordinate $X=(\xi _1+\xi _2)/2$, and relative coordinate $
x=\xi _2-\xi _1$, the value $\chi (\xi _1,\xi _2)$ can be written for $x<-a$
in the form 
\begin{equation}
\label{II.11}\exp (iq_1\xi _1+iq_2\xi _2)=\exp (iKX+ikx).
\end{equation}
If the incoming plane wave $\exp (ikx)$ is part of the solution, the
outgoing plane wave $\exp (-ikx)$ exists as well: 
\begin{equation}
\label{II.12}\exp (iKX-ikx)=\exp (iq_2\xi _1+iq_1\xi _2).
\end{equation}
These two waves differ from each other by permutation of the particle
quasi-momenta $q_1$ and $q_2$ only, and we do not get any additional
solutions apart from the class of solutions of Eq. (2.6). Given that the
expansion coefficients of the plane waves for $x<-a$ (in the region $\Gamma
_{-}$) are known, one can reconstruct the expansion coefficients for $x>a$
(in the region $\Gamma _{+}$). This is a standard problem in scattering
theory.

There exist two linearly independent solutions which can be taken to be
symmetric and antisymmetric under the substitution $x\leftrightarrow
-x\;(\xi _1\leftrightarrow \xi _2)$: 
\begin{equation}
\label{II.13}\chi _{_{+}}(\xi _1,\xi _2)=\exp (iKX)\left\{ 
\begin{array}{c}
e^{ikx}+S_{+}(k)e^{-ikx};\;
{\rm x<-a\;({region\;}\Gamma _{-})} \\ S_{+}(k)e^{ikx}+e^{-ikx};\;{\rm x>a\;(
{region\;}\Gamma _{+})}
\end{array}
\right. 
\end{equation}
\begin{equation}
\label{II.14}\chi _{_{-}}(\xi _1,\xi _2)=\exp (iKX)\left\{ 
\begin{array}{c}
e^{ikx}-S_{-}(k)e^{-ikx};\;
{\rm x<-a\;({region}}\;{\rm \Gamma _{-})} \\ 
S_{-}(k)e^{ikx}-e^{-ikx};\;{\rm x>a\;({region}\;\Gamma _{+}).}
\end{array}
\right. 
\end{equation}
The expansion coefficients in two neighboring regions of integrability are
related by the scattering matrices $S_{\pm }(k)$. The scattering problem can
be formulated on the half-axis $x\in (-\infty ,0]$ with the boundary
conditions $\chi _{_{+}}(X,0)^{^{\prime }}=0$ (symmetric case) and $\chi
_{_{+}}(X,0)=0$ (antisymmetric case), or, equivalently, on the half-axis $
[0,+\infty )$. The current density vanishes for $x=0$ for symmetric and
antisymmetric wave functions. From particle number conservation it follows
that the absolute values of the $S$-matrices are equal to unity on the real $
k$-axis. The following properties of the $S$ -matrix hold true in the whole
complex $k$-plane: 
$$
\begin{array}{c}
S_{\pm }(k)=S_{\pm }(-k)^{-1}, \\ 
S_{\pm }(k)=(S_{\pm }(k^{*})^{-1})^{*}.
\end{array}
$$
As usual, bound states are described by poles on the upper imaginary
half-axis, virtual levels are described by poles on the lower imaginary
half-axis. Poles in the lower half-plane of the complex $k$-plane
correspond to resonances. For wave functions of the general form, $\chi (X,x)
$ $=C_{+}\chi _{+}(X,x)+C_{-}\chi _{-}(X,x)$, the expansion coefficients for
incoming $e^{ikx}$ and outgoing $e^{-ikx}$ waves in the region $\Gamma _{-}$ 
$(x<-a)$ are connected unambiguously to the expansion coefficients in the
region $\Gamma _{+}$ $(x>a)$. The relation is expressed through the
scattering matrices $S_{\pm }(k)$ that can be obtained by solving equation
(2.10) on the interval $|\xi _2-\xi _1|<a$. We assume that this problem is
already solved and that the matrices $S_{\pm }(k)$ are known. In Eq. (2.13)
the signs of the $S_{\pm }(k)$ are fixed by the convention that for free
particles $S_{\pm }(k)=1$ and $\chi _{+}(X,x)\propto \cos (kx)$, $\chi
_{-}(X,x)\propto \sin (kx)$. Note that due to the boundary condition at $x=0$
, the value $\chi _{-}(x)$ $=\exp (-iKX)\chi _{-}(X,x)$ can be interpreted
as the radial part of the scattering wave function in three dimensions, and
the value $S_{-}(k)$, respectively, as the $S$-matrix corresponding to zero
angular momentum (the centrifugal potential for $l\neq 0$ cannot be included
in $V(x)$ since $V(x)$ is a short range potential).

Identifying the components of the wave function $\chi (X,x)$ $=C_{+}\chi
_{+}(X,x)+C_{-}\chi _{-}(X,x)$ with the relevant terms in the expansion of
Eq.(2.6), we obtain 
\begin{equation}
\label{II.15}
\begin{array}{c}
C_{\alpha _1...\alpha _s\alpha _r...\alpha _N}^{\gamma _1...\gamma _s\gamma
_r...\gamma _N}=C_{+}S_{+}+C_{-}S_{-}\; 
{\rm ({region\;}\Gamma _{+},{wave\ e}^{ikx}),} \\ C_{\alpha _1...\alpha
_r\alpha _s...\alpha _N}^{\gamma _1...\gamma _s\gamma _r...\gamma
_N}=C_{+}-C_{-}\;\;\;\;\;\; 
{\rm ({region\;}\Gamma _{+}{,wave\ e}^{-ikx}),} \\ C_{\alpha _1...\alpha
_r\alpha _s...\alpha _N}^{\gamma _1...\gamma _r\gamma _s...\gamma
_N}=C_{+}+C_{-}\;\;\;\;\;\; 
{\rm ({region\;}\Gamma _{-}{,wave\ e}^{ikx}),} \\ C_{\alpha _1...\alpha
_s\alpha _r...\alpha _N}^{\gamma _1...\gamma _r\gamma _s...\gamma
_N}=C_{+}S_{+}-C_{-}S_{-}\;{\rm ({region\;}\Gamma _{-}{,wave\ e}^{-ikx})} 
\end{array}
\end{equation}
where $S_{\pm }=S_{\pm }((k_{\alpha _r}-k_{\alpha _s})/2)$. This system of
equations is overdetermined. Yet, it admits a consistent solution for
matching the wave functions. From these equations we derive 
\begin{equation}
\label{II.16}C_{\alpha _1...\alpha _r\alpha _s...\alpha _N}^{\gamma
_1...\gamma _r\gamma _s...\gamma _N}=\frac{S_{-}-S_{+}}{S_{-}+S_{+}}
C_{\alpha _1...\alpha _r\alpha _s...\alpha _N}^{\gamma _1...\gamma _s\gamma
_r...\gamma _N}+\frac 2{S_{-}+S_{+}}C_{\alpha _1...\alpha _s\alpha
_r...\alpha _N}^{\gamma _1...\gamma _s\gamma _r...\gamma _N}. 
\end{equation}

Combining conditions (2.8) and (2.16), we obtain 
\begin{equation}
\label{II.17}C_{\alpha _1...\alpha _s\alpha _r...\alpha _N}^{\gamma
_1...\gamma _s\gamma _r...\gamma _N}=S_{+}((k_{\alpha _r}-k_{\alpha
_s})/2)C_{\alpha _1...\alpha _r\alpha _s...\alpha _N}^{\gamma _1...\gamma
_s\gamma _r...\gamma _N}. 
\end{equation}
If we make one more permutation of the indices $\alpha _s$, $\alpha _r$, we
obtain, by virtue of $S_{+}((k_{\alpha _r}-k_{\alpha _s})/2)$ $
S_{+}((k_{\alpha _s}-k_{\alpha _r})/2)=1$, the initial expression which
completes this consistency check.

\section{GENERALIZED BOHR-SOMMERFELD\protect\\ QUANTIZATION RULES}

\setcounter{equation}{0}

It follows from condition (2.17) that the expansion coefficients are all
expressed unambiguously through $C_{1...N}^{1...N}$. Using Eq.(2.8), the
upper indices can be ordered in sequence $(1,...,N)$. The expansion
coefficients, which can be obtained from each other by permutation of two
neighboring indices, are related by Eq.(2.17). The arbitrary set of indices $
(\alpha _1,...,\alpha _N)$ can be obtained from the sequence $(1,...,N)$ by
permutations of neighboring indices. Thus, one can express $C_{\alpha
_1...\alpha _N}^{1...N}$ through $C_{1...N}^{1...N}$. It remains to verify
that different sequences of transpositions transforming $(1,...,N)$ to 
$(\alpha _1,...,\alpha _N)$ 
give the same result. Let $P_1$ and $P_2$ be two
such permutations of the initial sequence $(1,....,N)$ leading to the same
final sequence  $(\alpha _1,...,\alpha _N)$.
It is evident that the permutation $P_1\times P_2^{-1}$
describes an identical (trivial) transformation. In such a sequence, each
pair of indices changes places an even number of times. Since $
S_{+}((k_{\alpha _r}-k_{\alpha _s})/2)$ $S_{+}((k_{\alpha _s}-k_{\alpha
_r})/2)=1$, the result of the transformation $P_1\times P_2^{-1}$ is the
identity $C_{1...N}^{1...N}=C_{1...N}^{1...N}$. Denoting the result\- of
transformation $P_1$ by 
\begin{equation}
\label{III.1}C_{\alpha _1...\alpha _N}^{1...N}=\alpha C_{1...N}^{1...N}
\end{equation}
and the result of transformation $P_2^{-1}$ by 
\begin{equation}
\label{III.2}C_{1...N}^{1...N}=\beta C_{\alpha _1...\alpha _N}^{1...N}
\end{equation}
we obtain $\alpha \beta =1$. Therefore, from both permutations $P_1$ and $P_2
$ we obtain one and the same relation (3.1). The phase shift $\delta _{+}(k)$
is defined by equation $S_{+}(k)=\exp (2i\delta _{+}(k))$. With the help of
eqs. (2.7), (2.8) and (2.17) we obtain the generalized Bohr-Sommerfeld
quantization rule 
\begin{equation}
\label{III.3}k_jL+\sum_{l=1}^N2\delta _{+}(\frac{k_j-k_l}2)=2\pi n_j.
\end{equation}

The sum is running over $l\neq j,$ $n_j$ are integer numbers. This equation
can be interpreted in the following way. Going around the circle $[0,L]$,
the particle $j$ scatters on each particle $l\neq j$, acquiring an
additional phase $2\delta _{+}(\frac{k_j-k_l}2)$. When the particle $j$ is
back to the initial place, its phase turns out to be equal to the left hand
side of Eq.(\ref{III.3}), the additional term $k_jL$ is a result of the
translation. The total phase must be an integer multiple of $2\pi $, since
the wave function is single valued.

Let us now consider the case when one virtual level exists in the complex $k$
-plane. In this case the $S$-matrix 
\begin{equation}
\label{III.4}S_{+}(k)=(k-ik_0)/(k+ik_0),
\end{equation}
where $k_0=m\alpha >0$, corresponds to the 
delta-function potential $
V(x)=\alpha \delta (x)$ (problem 2.47 in ref. \cite{Gal}). Respectively, $
\delta _{+}(-\infty )=0$, $\delta _{+}(+\infty )=2\pi $, $\delta _{+}(0)=\pi 
$, so that $S_{+}(0)=-1$. It is seen from eqs.(2.13) that the symmetric wave
function vanishes for  $k=0$ and $S_{+}(0)=-1$ and the particle quasi-momenta
in the set $(k_1,k_2,...,k_N)$ must be all distinct. Therefore, one obtains
a constraint which resembles the Pauli principle even though we have started
with a system of bosons. In the lowest energy state, the particle
quasi-momenta $(k_1,k_2,...,k_N)$ occupy the Fermi sphere. The equality $
S_{+}(0)=-1$, along with the exclusion principle, apparently, is valid for
any odd number of virtual levels.

For an even (zero) number of virtual levels one has $S_{+}(0)=1$, and some
quasi-momenta in the set $(k_1,k_2,...,k_N)$ can coincide. However, for $
q_1=q_2$, $|x|>a$, Eq. (2.10) has the general solution $\chi
_{+}(X,x)=(a+bx)\exp (iKX)$ with $b\neq 0$. For this reason it is impossible
to satisfy the periodic boundary condition. The states with $b\neq 0$
correspond to zero-energy scattering states. The coefficient $b$ vanishes
for a special class of potentials having a representation of the form 
\begin{equation}
\label{III.5}V(x)=\frac 1m\chi ^{\prime \prime }(x)/\chi (x)
\end{equation}
with the wave function satisfying the condition $\chi ^{\prime }(0)=\chi
^{\prime }(a)=0$. From $V(a)=0$ it also follows that $\chi ^{\prime \prime
}(a)=0$. Solutions of the Schr\"odinger equation with a limited asymptotic
behavior at infinity (i.e. $b=0$) exist when a discrete level in the
potential appears (problem 2.18 in ref. \cite{Gal}). Therefore, it is clear
that the case $b=0$ is an exceptional one. Solutions of such a kind occur
when we pass to a system in which bound states are formed.

We thus consider repulsive potentials without discrete states in the energy
spectrum. For such potentials all quasi-momenta in the set $
(k_1,k_2,...,k_N) $ must be distinct.

\section{THERMODYNAMIC LIMIT}

\setcounter{equation}{0}

In the thermodynamic limit $N\rightarrow \infty $ , $L\rightarrow \infty $ ,
and $\rho =N/L$ a fixed value, the particles in the ground state occupy
continuously the Fermi sphere. The sum over the particle quasi-momenta can
be approximated by an integral over $k$ with a weight function $f(k)$ 
$$
\sum \rightarrow \int \frac{Ldk}{2\pi }f(k), 
$$
where $Lf(k)dk/(2\pi )$ is the number of states in the interval $dk$. The
distribution function $f(k)$ can be found from equation 
\begin{equation}
\label{IV.1}f(k)=1+\int_{-p_F}^{p_F}\frac{dk^{\prime }}{2\pi }f(k^{\prime
})\delta _{+}^{\prime }(\frac{k-k^{\prime }}2)
\end{equation}
where $\delta _{+}^{\prime }(\frac{k-k^{\prime }}2)$ is the derivative of
the scattering phase shift with respect to the argument. This equation is
obtained by rewriting Eq.(3.3) in the thermodynamic limit for the phase
difference of the particles $j+1$ and $j$ taking into account that $
n_{j+1}-n_j=1$. Eq.(4.1) is a generalization of Eq.(3.12) in ref. [3] to
arbitrary singular potentials.

We consider the problem of finding the spectrum of elementary excitations.
Let us consider two sets of particle quasi-momenta $(k_1,k_2,...,k_N)$ and $
(k_1^{\prime },k_2^{\prime },...,k_N^{\prime })$. In the first set the $k$'s
occupy continuously the Fermi sphere. The second set of $k$'s is obtained
from the first one by removing a particle from the Fermi surface with the
quasi-momentum $k_N=p_F$ and giving it an arbitrary quasi-momentum $
q=k_N^{\prime }>p_F$ or $q=k_N^{\prime }<-p_F$. Because of the interaction,
the particle quasi-momenta inside the Fermi sphere receive a shift $
k_j^{\prime }-k_j=\Delta (k_j)/L$. Taking the difference between the two
relations 
$$
k_jL+\sum_{l=1}^{N-1}2\delta _{+}(\frac{k_j-k_l}2)+2\delta _{+}(\frac{k_j-p_F
}2)=2\pi n_j, 
$$
$$
k_j^{\prime }L+\sum_{l=1}^{N-1}2\delta _{+}(\frac{k_j^{\prime }-k_l^{\prime }
}2)+2\delta _{+}(\frac{k_j^{\prime }-p_F}2)=2\pi n_j, 
$$
we obtain with the help of Eq.(4.1), 
\begin{equation}
\label{IV.2}\Delta (k)f(k)=-2\delta _{+}(\frac{k-q}2)+2\delta _{+}(\frac{
k-p_F}2)+\int_{-p_F}^{p_F}\frac{dk^{\prime }}{2\pi }\Delta (k^{\prime
})f(k^{\prime })\delta _{+}^{\prime }(\frac{k-k^{\prime }}2).
\end{equation}
The total momentum of the system equals 
\begin{equation}
\label{IV.3}P=q-p_F+\int_{-p_F}^{p_F}\frac{dk^{\prime }}{2\pi }\Delta
(k^{\prime })f(k^{\prime }).
\end{equation}
The first term is the quasi-momentum of the excited particle. After removing
the particle with quasi-momentum $p_F$ the momentum of the Fermi sphere
changes by an amount $-p_F$. The last term represents the change of the
total momentum of the particles inside the Fermi sphere. The energy of the
system equals 
\begin{equation}
\label{IV.4}\epsilon =\frac{q^2}{2m}-\frac{p_F^2}{2m}+\int_{-p_F}^{p_F}\frac{
dk^{\prime }}{2\pi }\frac{k^{\prime }}m\Delta (k^{\prime })f(k^{\prime }).
\end{equation}
With $q\rightarrow p_F$ the momentum $P$ and the energy $\epsilon $ of the
excited state vanish. In order to find the dispersion law for elementary
excitations, $\epsilon (P)$, it is necessary to express $q$ in terms of $P$
with the help of Eq.(4.3) and substitute the resulting expression into
Eq.(4.4). Note that the integral equations (4.1) and (4.2) have the same
kernel 
\begin{equation}
\label{IV.5}R(k,k^{\prime })=(2\pi )\delta (k-k^{\prime })-\delta
_{+}^{\prime }(\frac{k-k^{\prime }}2)
\end{equation}
and that $R(k,k^{\prime })$ is symmetric. Solutions of these equations can
be represented in the form 
\begin{equation}
\label{IV.6}f(k)=\int_{-p_F}^{p_F}\frac{dk^{\prime }}{2\pi }
R^{-1}(k,k^{\prime }),
\end{equation}
\begin{equation}
\label{IV.7}\Delta (k)f(k)=\int_{-p_F}^{p_F}\frac{dk^{\prime }}{2\pi }
R^{-1}(k,k^{\prime })(-2\delta _{+}(\frac{k^{\prime }-q}2)+2\delta _{+}(
\frac{k^{\prime }-p_F}2)).
\end{equation}
Integrating Eq.(\ref{IV.7}) over $k$ and using the symmetry under
permutation of the arguments of $R^{-1}(k,k^{\prime })$, we can rewrite
Eq.(4.3) in the form 
\begin{equation}
\label{IV.8}P=q-p_F+\int_{-p_F}^{p_F}\frac{dk^{\prime }}{2\pi }f(k^{\prime
})(-2\delta _{+}(\frac{k^{\prime }-q}2)+2\delta _{+}(\frac{k^{\prime }-p_F}
2)).
\end{equation}

In lowest order of the difference $q-p_F$ 
\begin{equation}
\label{IV.9}P=(q-p_F)f(p_F), 
\end{equation}

\begin{equation}
\label{IV.10}\epsilon =(q-p_F)\int_{-p_F}^{p_F}\frac{dk^{\prime }}{2\pi } 
\frac{k^{\prime }}mR^{-1}(k^{\prime },p_F). 
\end{equation}

In a similar way one can treat excitations which transfer a particle to the
opposite side of the Fermi surface, that is for $k_1=-p_F$ and $
q=k_1^{\prime }>p_F$ or $q=k_1^{\prime }<-p_F$. These excitations are
described by the same formulae, if the substitution $p_F\rightarrow -p_F$ is
made.

Other kinds of elementary excitations of ''hole'' type (non-Bogoliubov
excitations) are obtained by taking away a particle with quasi-momentum $q$
inside of the Fermi sphere and placing it on the Fermi surface. Let us find
the spectrum of such excitations. Consider two sets of the particle
quasi-momenta $(k_1,k_2,...,k_N)$ and $(k_1^{\prime },k_2^{\prime
},...,k_N^{\prime })$. In the first set the quasi-momenta occupy
continuously the Fermi sphere, while the second set is obtained from the
first one by removing a particle with a quasi-momentum $k_m=q$ from the
Fermi sphere and placing it on the Fermi surface at $k_m^{\prime }=p_F$. By
comparison of the relations 
$$
k_jL+\sum_{l=1}^{N-1}2\delta _{+}(\frac{k_j-k_l}2)+2\delta _{+}(\frac{k_j-q}
2)=2n_j, 
$$

$$
k_j^{\prime }L+\sum_{l=1}^{N-1}2\delta _{+}(\frac{k_j^{\prime }-k_l^{\prime }
}2)+2\delta _{+}(\frac{k_j^{\prime }-p_F}2)=2n_j 
$$
with the corresponding relations for ''particle'' type excitations,
discussed before, we see that the corresponding equations can be obtained
from the ones already discussed by the replacement $p_F\rightarrow q$. In
the above formulae the terms $l=m$ are excluded from the summation.

In the symmetric case, when the particle with quasi-momentum $k_m=q$ is
placed on the Fermi surface at $k_m^{\prime }=-p_F$, it is necessary to make
the replacement $p_F\rightarrow -p_F$.

\section{EXACTLY SOLVABLE MODELS FOR FERMI-SYSTEMS}

\setcounter{equation}{0}

In full analogy to the Bose case, the problem of constructing the wave
function of the Fermi-system can be solved when the spins of the fermions
are all lined up in one direction. In such a case, the spin part of the wave
function is symmetric, and the coordinate part is totally antisymmetric. Let
us find how eqs.(2.7), (2.8), (2.17) and (3.3) should be modified.

The periodic boundary condition (2.7), apparently, remains unaltered.

Under permutations of the coordinates, the wave function is antisymmetric.
The symmetry conditions yield the expansion coefficients satisfying the
condition: 
\begin{equation}
\label{V.1}C_{\alpha _1...\alpha _s...\alpha _r...\alpha _N}^{\gamma
_1...\gamma _s...\gamma _r...\gamma _N}=-C_{\alpha _1...\alpha _s...\alpha
_r...\alpha _N}^{\gamma _1...\gamma _r...\gamma _s...\gamma _N}. 
\end{equation}
In comparison to the Bose case, the right hand side acquires a minus sign.

The matching conditions of the wave function at the boundaries of the
different regions of integrability are derived as for bosons. Relations
(2.16) are valid. Combining eqs.(2.16) and (5.1), we obtain 
\begin{equation}
\label{V.2}C_{\alpha _1...\alpha _s\alpha _r...\alpha _N}^{\gamma
_1...\gamma _s\gamma _r...\gamma _N}=S_{-}((k_{\alpha _r}-k_{\alpha
_s})/2)C_{\alpha _1...\alpha _r\alpha _s...\alpha _N}^{\gamma _1...\gamma
_s\gamma _r...\gamma _N}.
\end{equation}

The expansion coefficients are related to each other through the matrix $
S_{-}(k)$ that describes scattering on the positive half-axis with the
boundary condition $\chi (0)=0$.

The generalized Bohr-Sommerfeld quantization rules have a form analogous to
Eq.(3.3)

\begin{equation}
\label{V.3}k_jL+\sum_{l=1}^N2\delta _{-}(\frac{k_j-k_l}2)=2\pi n_j. 
\end{equation}
where the phase shift is defined by $S_{-}(k)=\exp (2i\delta _{-}(k))$. The
summation is performed over $l\neq j$, $n_j$ are integer numbers.

Therefore, there is a close analogy between the the behavior of Bose- and
Fermi-systems with parallel spins. This analogy also exists in the
thermodynamic limit. The results of Sect.4 are valid for Fermi-systems after
the replacement $\delta _{+}(k)\leftrightarrow \delta _{-}(k)$.

In the limit $k\rightarrow 0$, the scattering phase shifts $\delta _{+}(k)$
and $\delta _{-}(k)$ have a different behavior. For smooth potentials, the
continuity conditions for the logarithmic derivatives at $x=a$ in the
symmetric and antisymmetric cases have the form 
\begin{equation}
\label{V.4}k\tan (ka+\delta _{+}(k))=\kappa _{+},
\end{equation}
\begin{equation}
\label{V.5}k\cot (ka+\delta _{-}(k))=\kappa _{-},
\end{equation}
$k_{+}$ and $k_{-}$ do not depend on the momentum $k$ if $k<<\kappa _{+},$ $
\kappa _{-}$. We conclude that at small $k$ the scattering phase shifts have
the form 
\begin{equation}
\label{V.6}\delta _{+}(k)=-ka+\arctan (\kappa _{+}/k),
\end{equation}
\begin{equation}
\label{V.7}\delta _{-}(k)=-ka+{\rm arccot}(\kappa _{-}/k).
\end{equation}
In the limit $a\rightarrow 0$ and for arbitrary small but finite values of $
\kappa _{+}$ and $\kappa _{-}$, we obtain 
\begin{equation}
\label{V.8}\delta _{+}(k)\rightarrow \pi \theta (k),
\end{equation}
\begin{equation}
\label{V.9}\delta _{-}(k)\rightarrow 0.
\end{equation}
In the weak coupling regime (the value of the $\kappa _{+}$ is small) the
quantization conditions (2.17) for the symmetric case remain non-trivial,
since the derivative of $\delta _{+}(k)$ is proportional to a
delta-function. The integral kernel $R(k,k^{\prime })$ defined by Eq.(4.5)
is determined by the difference of the delta-function and the derivative of $
\delta _{+}(k)$. For Bose-systems with the potential $V(x)=\alpha \delta (x)$
, it is impossible to reproduce the result of perturbation theory
analytically beyond first order of $\alpha $ [4], because of the complicated
character of the kernel $R(k,k^{\prime })$. In general it is, however, not
difficult to perform a comparison with perturbation theory numerically. The
non-trivial character of the weak coupling regime can be related to the
non-analyticity in the coupling constant (in $\lambda $ if we write the
potential as $\lambda V(x)$), that should be present because of the
instability of the system of bosons when the sign of the potential is
changed.

According to Eq.(5.9) the phase $\delta _{-}(k)$ and its derivative should
be set equal to zero. After that, the quantization conditions (5.3) take a
very simple form $k_jL=2\pi n_j$. As a result, we deal with a Fermi-gas of
non-interacting particles. The weak coupling regime is therefore trivial for
fermions.

\section{BOSE-EINSTEIN CONDENSATION OF TWO-FERMION RESONANCES}

\setcounter{equation}{0}

In this sect. we consider a problem of considerable physical interest. As in
sect. 5 consider a one-dimensional system of fermions interacting via a
finite ranged potential $V(x)$. Suppose there exists in the fermion-fermion
channel a resonance with momentum $k_0=k_1-ik_2$, where $k_1$ and $k_2$ are
real, positive numbers. One expects that after increasing the Fermi momentum $
k_F$ above the value of the resonance $k_1$, the creation of such resonances
will be energetically favored by the system. The resonances can be treated
as composite Bose particles. If their interaction is small, they are
accumulated in the ground state like real bosons with zero total momentum.
The Bose-Einstein condensation reveals itself by an increase of the
distribution function $f(k)$ of the particle quasi-momenta near the Fermi
surface, since the resonances are at rest and the fermion momenta are
concentrated in the vicinity of $k\approx k_1\approx k_F$. The effect is
well pronounced for a small width and disappears with increasing width of
the resonance.

Near the resonance, the $S$-matrix can be parametrized in the Breit-Wigner
form 
\begin{equation}
\label{VI.1}S_{-}(k)=\frac{(k+k_0)(k-k_0^{*})}{(k+k_0^{*})(k-k_0)}, 
\end{equation}

An $S$-matrix of such a type corresponds not only to finite range
potentials, but also to zero-range singular potentials.

Let us consider a potential of the form 
\begin{equation}
\label{VI.2}V(x)=-V_0\theta (a-x)+\alpha \delta (x-a), 
\end{equation}
where $V_0$ represents a positive value. We wish to find a solution of the
scattering problem on the half-axis $[0,+\infty )$. The wave function takes
the form 
\begin{equation}
\label{VI.3}\chi (x)=\left\{ 
\begin{array}{c}
{\sin (Kx);\;\;\;\;\;0<x<a,} \\ {C{\sin }(kx+\delta }_{{-}}{(k));\;a<x}. 
\end{array}
\right. 
\end{equation}
Here, $K=\sqrt{k^2+V_0}$ (in ''atomic units'' $m=\hbar =1$). The scattering
phase is determined by the condition 
\begin{equation}
\label{VI.4}k\cot (ka+\delta _{-}(k))=K\cot (Ka)+\alpha . 
\end{equation}
The $S$-matrix takes the form 
\begin{equation}
\label{VI.5}S_{-}(k)=\exp (-2ika)\frac{K\cot (Ka)+\alpha +ik}{K\cot
(Ka)+\alpha -ik}. 
\end{equation}
The $S$-matrix poles can be found from the equation 
\begin{equation}
\label{VI.6}\tan (Ka)=\frac K{ik-\alpha }. 
\end{equation}
In the limit $\alpha \rightarrow \infty $ , the $S$-matrix poles cross the
real axis at $Ka=0$ ($mod$ $\pi $ ). The lowest level is located at $Ka=\pi $
. We may keep the lowest resonance momentum $k_1=\sqrt{\pi ^2/a^2-V_0}$
fixed and pass to the limit $V_0\rightarrow \infty \ $and $a\rightarrow 0.$
The zero-width resonance occurs then from a solution to the Schr\"odinger
equation with the zero-range singular potential (6.2). The potential is
determined by the limiting procedure: $V_0\rightarrow \infty $, $
a\rightarrow 0$, $k_1=\sqrt{\pi ^2/a^2-V_0}=constant$. It generates on the
real $k$-axis exactly two poles (one zero-width resonance) in the $S$
-matrix. The other poles are moved to infinity.

Let now $\alpha $ be finite, but large, and $V_0$ receives a correction $
\Delta V_0<<V_0$. The resonance acquires, first of all, a finite width. We
require the $\Delta V_0$ be such that the real part of the resonance
momentum be equal to $k_1$. The $S$-matrix pole is located at $k_0=k_1-ik_2$.

Now, we parametrize $\alpha =1/(\xi x^3)$ and $a=x^2$, so that 
\begin{equation}
\label{VI.7}V_0/\pi ^2=\frac 1{x^4}-k_1^2. 
\end{equation}
In the limit $x\rightarrow 0$, Eq.(6.6) gives 
\begin{equation}
\label{VI.8}\Delta V_0/\pi ^2=-\frac{2\xi }{x^3}+\frac{3\xi ^2}{x^2}-\frac{
2\xi ^3}{3x}(6-\pi ^2-3x^4k_1^2)+\frac{\xi ^4}3(10-13\pi
^2-27x^4k_1^2)+...\;, 
\end{equation}
\begin{equation}
\label{VI.9}k_2/\pi ^2=\xi ^2-3x\xi ^3+(6-\pi ^2-x^4k_1^2)x^2\xi ^4+...\;. 
\end{equation}
As a result, we obtain a zero-range singular potential for which the $S$
-matrix in the complex $k$-plane has two poles only, which can be identified
with a resonance. The resonance width is proportional to $\xi ^2.$ For
narrow resonances $\xi <<1.$

In Fig.1, we show results for the distribution function $f(k)$ for different
values of the Fermi momentum $k_F$ from $0.64$ to $1.6$ with a step $0.12$
in a system where the two-fermion interaction is described by a $S$-matrix
of the form (6.1) with $k_1=1$, $k_2=0.05$ (small width). 
The distribution function $f(k)$
increases near the Fermi surface where the production of the resonances
becomes energetically favorable. This effect can be interpreted in terms of
a Bose-Einstein condensation of the resonances whose wave function are
concentrated at $k\approx k_1\approx k_F$. The value $k_2$ measures the
spread of the fermion momenta in the resonances. When the Fermi momentum is
increased beyond the resonance momentum $k_1$ the distribution function
shows a plateau centered at $k\approx k_1$. If all resonances are in the
Bose-Einstein condensate, the size of the plateau would be of order $k_2$.
However, its size increases with the Fermi momentum. This can be interpreted
as follows: It is known that in systems of interacting bosons there exists a
fraction of bosons with nonvanishing velocities which are not in the
condensate and which have a nontrivial momentum distribution \cite{Lif}.
These bosons contribute to the pressure. Their fraction increases with the
total density of the bosons.

The appearance of resonances in Fermi systems yields a softer equation of
state. This means that the pressure $p$ increases slower with the density $n$
. The effect is shown in Fig. 2. The sudden rise of $p=p(n)$ appears at $
k_F=k_1$ where the particle density equals $n=0.36$.

For a broad resonance, the derivative of the scattering phase $\delta
^{^{\prime }}(\frac{k-k^{\prime }}2)$ entering Eq.(4.2) is small, the
distribution function $f(k)$ is close to unity, and therfore the effect of the
resonance on the equation of state is small, too.

\section{SELF-GRAVITATING OBJECTS}

\setcounter{equation}{0}

Bose-Einstein condensation of narrow two-fermion resonances may drastically
change the properties of fermion matter, producing physically interesting
phenomena, for example, the instability of neutron stars. Before, studying
this effect in more detail let us qualitatively describe the expected
scenario.

Narrow two-fermion resonances can be treated as Bose particles. If the
central density of a neutron star exceeds a critical value, creation of
these bosons with subsequent formation of a Bose-Einstein condensate becomes
energetically favorable. The critical density for a Bose-Einstein condensate
formation is determined by the mass of the resonances. In the ideal Bose-gas
approximation, the chemical potential of the fermions is frozen at $\mu
=m_D/2$, where $m_D$ is the dibaryon mass. The dibaryons are at rest.
Therefore, they do not collide with the boundary and do not contribute to
the pressure. The pressure is determined by the fermions only. The number of
fermions remains constant with increasing density, since the radius of the
Fermi sphere is frozen, whereas the number of the resonances increases
linearly. The incompressibility of matter vanishes. In this way, we give a
qualitative explanation for the observed growth of the distribution function 
$f(k)$ near the Fermi surface in Fig. 1 as well as the rapid change of $
p=p(n)$ in Fig. 2.

Suppose that the short-range potential between fermions is such that a
narrow two-fermion (dibaryon) resonance exists and that a Bose-Einstein
condensate of these resonances is formed in the interior of a neutron star
(in three dimensions) for $r<r_1$. In the inner region, due to the formation
of the resonances, the pressure remains constant, i.e. ${\bf \nabla }p=0$.
Gauss's law implies 
$$
\int d{\bf S \cdot \nabla }\phi (r)=4\pi GM(r) 
$$
where $M(r)$ is mass of the substance inside of a sphere of the radius $r$
and $G$ the gravitational constant. We conclude that ${\bf \nabla }\phi
(r)\neq 0$. The Euler equation 
\begin{equation}
\label{VII.1}\rho \frac{\partial {\bf v}}{\partial t}+\rho ({\bf v}\cdot 
{\bf \nabla }){\bf v}=-{\bf \nabla }p-\rho {\bf \nabla }\phi (r) 
\end{equation}
shows that in the static case (${\bf v}\equiv 0$) the gradient of the
pressure $p$ is balanced by the gradient of the gravitational potential $
\phi (r)$. However, if ${\bf \nabla }p=0$, static solutions are impossible
and neutron stars are gravitationally unstable.

Let us study this effect in a one-dimensional model of the neutron star,
where all quantities can be calculated exactly, in order to test if some
instability occurs due to the formation of the two-fermion resonances. We
restrict ourselves to nonrelativistic Newtonian gravity. The static stable
configurations of neutron stars are described by Euler's equation with a
vanishing left hand side 
\begin{equation}
\label{VII.2}-\frac{dp(x)}{dx}-\rho (x)\frac{d\phi (x)}{dx}=0
\end{equation}
where $p(x)$ is the pressure, $\rho (x)=mn(x)$ the mass density, $n(x)$ the
number density, and $\phi (x)$ the gravitational potential. The center of
the neutron star is placed at the origin of the coordinates. Gauss's law
gives 
\begin{equation}
\label{VII.3}\frac{d^2\phi (x)}{dx^2}=4\pi G\rho (x).
\end{equation}
Note that in contrast to classical electrodynamics the right hand side of
Eq.(\ref{VII.3}) has a positive sign.

We integrate Eq.(7.3) from $-x$ to $+x$. Using the symmetry of the potential 
$\phi (x)$ under the reflection $x\leftrightarrow -x$ ($p(x)$ and $\rho (x)$
are also symmetric functions) and the antisymmetry of the first derivative
of $\phi (x)$, one gets 
\begin{equation}
\label{VII.4}\frac{d\phi (x)}{dx}=2\pi GM(x), 
\end{equation}
where 
\begin{equation}
\label{VII.5}M(x)=\int_{-x}^xdx\rho (x)=2\int_0^xdx\rho (x). 
\end{equation}
Substituting Eq.(7.4) into Eq.(7.2), we obtain 
\begin{equation}
\label{VII.6}\frac{dp}{dM}+\pi GM=0, 
\end{equation}
and finally 
\begin{equation}
\label{VII.7}M(x)=\sqrt{\frac 2{\pi G}(p(0)-p(x))}. 
\end{equation}
At the surface $p(x_s)=0$, and the total mass of the neutron star $
M_s=M(x_s) $ is expressed unambiguously through the central pressure $p(0)$.
The neutron star radius $x_s$ can be obtained from Eq.(7.7). It is
sufficient to take the derivative of Eq.(7.7) with respect to $x$, divide
both sides by $2\rho (x)$, and integrate the result over $x$. In this way
one gets 
\begin{equation}
\label{VII.8}x_s=\frac 1{\sqrt{2\pi G}}\int_0^{p(0)}\frac{dp}{\rho (p)\sqrt{
p(0)-p}} 
\end{equation}

Assuming that the equation of state $n=n(p)$ is known, eqs.(7.7) and (7.8)
allow to determine the neutron star radius $x_s$ as a function of 
the neutron star
mass $M_s$, or, equivalently, the total mass $M_s$ as a function of the central
pressure $p(0)$. The criterion for the gravitational stability of stars has
the form (see ref. \cite{Zel}, Eq.(10.1.4p)) 
\begin{equation}
\label{VII.9}\frac{\partial x(M)}{\partial M_s}<0,
\end{equation}
where $x(M)$ is a coordinate, such that inside of the interval $[-x,x]$ the
matter of mass $M$ is contained: 
\begin{equation}
\label{VII.10}x(M)=\int_0^M\frac{dM^{\prime }}{2\rho }.
\end{equation}
Taking the derivative with respect to $M_s$ and using Eq. (7.7) to express
the pressure in terms of the mass, we obtain 
\begin{equation}
\label{VII.11}\frac{\partial x(M)}{\partial M_s}=-\frac{\pi GM_s}2\int_0^M
\frac{dM^{\prime }}{\rho ^2}\frac{d\rho }{dp}<0.
\end{equation}
In the non-degenerate case $dp/d\rho =a_s^2>0$ ($a_s$ is the velocity of
sound), the derivative ${\partial x(M)}/{\partial M_s}$ is negative definite
and one-dimensional neutron stars are in general gravitationally stable.

In three-dimensional space this is not necessarily the case. In order to
investigate the reason for this difference, let us express the sum of the
internal energy and the gravitational binding energy 
\begin{equation}
\label{VII.12}E=\int \epsilon dV+{\rm {sign}}(2-D)G\int r^{2-D}MdM 
\end{equation}
in terms of the average density $\bar \rho \sim ~M/r^D$ where $D$ is the
dimension of the space. For an equation of state of the form $\varepsilon
=A\rho ^\gamma $ we get 
\begin{equation}
\label{VII.13}E=A\bar \rho ^{\gamma -1}M+{\rm sign}(2-D)BM^{(2+D)/D}\bar
\rho ^{(D-2)/D} 
\end{equation}
where $B$ is a fixed positive constant.

In the one-dimensional case $D=1$, a minimum energy $E$ as a function of $
\bar \rho $ is obtained for $\gamma >1$. In the ideal gas approximation for
the Bose-Einstein condensate $\gamma =1$, $\rho \sim ~m_Dn$, where $n$ is
the number density, and so there is no stability. The maximum pressure
determines the maximum mass of the neutron star 
\begin{equation}
\label{VII.14}M_{\max }=\sqrt{\frac 2{\pi G}p_{\max }(0)}. 
\end{equation}
However, if one goes beyond the ideal gas approximation, after the creation
of the resonances the pressure still increases slowly with the density due
to the interactions between the fermions and bosons as shown in Fig. 2. In
the one-dimensional problem, even an arbitrarily slow growth of the pressure
($\gamma =1+0.0...01$) stabilizes the neutron star.

In the three-dimensional case, the minimum of $E$ exists for $\gamma >4/3$,
and for the stabilization of the neutron star, an arbitrarily slow growth of
pressure is insufficient. In order to make the neutron star stable, it is
necessary to increase the average value of $\gamma $ above $4/3$ (see \cite
{Zel}). Consequently, in the presence of a dibaryon condensate a
considerably stiffer equation of state is required to prevent the neutron
star from collapsing.

\section{CONCLUSION AND DISCUSSIONS}

\setcounter{equation}{0}

We have shown that the properties of one-dimensional Bose- and some
Fermi-systems can be determined exactly without recourse to perturbation
theory for a wide class of singular potentials. The solutions are
expressible through two-particle scattering phase shifts. The Fermi
character of the energy spectrum in one-dimensional Bose-systems is not
specific to potentials of delta-function type. In any singular
potential there are hole-type non-Bogoliubov branches of elementary
excitations. We have derived the dispersion laws for these excitations.

The one-dimensional Fermi system can be analyzed exactly if the fermion
spins are lined up in one direction. The quantization rules have a similar
form for Bose- and Fermi-systems.

An exact analysis of Fermi-systems with a resonance in the two-fermion
channel is given. In the model considered, the Pauli principle and the
composite nature of the resonances are taken into account at the outset. We
have observed an increase of the distribution function $f(k)$ of fermions
over the quasi-momenta $k$ in the vicinity of the Fermi surface when the
density is close to the critical density of resonance formation. We could
interpret this behavior in terms of a Bose-Einstein condensation of
two-fermion resonances. The formation of the resonances is accompanied by a
softening of the equation of state. In the real (three-dimensional) world
softening of the equation of state of nuclear matter caused by dibaryon
resonances can produce instability of neutron stars. In a one-dimensional
model of self-gravitating objects this effect does not exist.

The existence of a dibaryon resonance will lead at higher densities to the
occurrence of a new state of nuclear matter. The dibaryons behave like
bosons and can form a Bose-Einstein condensate in nuclear matter. With
increasing baryon density, the pressure should only slightly increase with
the density. In a perturbative picture (which was not used here), the
dibaryon condensation should occur at a Fermi energy $\varepsilon _F$
determined by the condition 
$$
2(m_N+\varepsilon _F)\geq m_D 
$$
where $m_N$ and $m_D$ are the nucleon and dibaryon masses.

It will be interesting to search for such a dibaryon condensate in heavy-ion
collisions. In the center-of-mass frame of the condensate a large fraction
of dibaryons has zero velocities. When the density decreases, dibaryons in
the condensate decay into nucleons. If a channel $D\rightarrow NN$ exists,
experimentalists will observe monochromatic nucleons with the energy $m_D/2$
. Since the rest frame of the condensate is {\it a priori} unknown, it is
necessary to look for a boost transformation along the beam momentum into a
coordinate system in which a statistically significant excess of
monochromatic nucleons exists. An excess of such events can be considered as a
possible signature for the formation of a dibaryon condensate.

It is also interesting to check astrophysical data for the presence of a
dibaryon condensate in the interiors of massive neutron stars as well as for
signatures of instability of neutron stars caused by dibaryons.

\vspace{1 cm}

{\bf ACKNOWLEDGMENTS}

We thank H.Frahm and B.V.Martemyanov for very useful discussions. One of the
authors (M.I.K.) wishes to thank Alexander von Humboldt-Stiftung for a
Forschungsstipendium and Neveu-INTAS and INTAS for financial support by
Grants No. 93-0023 and 93-0079.

\newpage

\centerline{FIGURE CAPTIONS}

\noindent
{\bf Fig.1} 
The distribution function $f(k)$ versus the quasi-momentum $k<k_F$
for 8 different Fermi momenta $k_F$ ranging from $0.64$ to $1.6$ with a step 
$0.12$ (in units $m=\hbar =1$). The two-body $S$-matrix has a pole
corresponding to a resonance at $k=k_1-ik_2$ with $k_1=1$ and  
$k_2=0.05$ (i.e. small width). 
The distribution function $f(k)$ increases near the Fermi surface when $k_F$
approaches $k_1$. This effect can be interpreted in terms of a Bose-Einstein
condensation of the resonances, since the wave function of the resonance is
concentrated at $k\approx k_1$. \\

\vspace{2.0cm}

\noindent
{\bf Fig.2} Pressure $p$ versus fermion number density $n$. The change of
the slope occurs at $n=0.36$ when $k_F=k_1$ (in units $m=\hbar =1$). This is
the point at which the two-fermion condensation starts. It results in a
softening of the equation of state. In the ideal gas approximation, the
pressure does not increase with the density after the condensation has set
in. If interactions are included, the pressure is slowly increasing even
after the condensation.

\newpage

\begin{figure}[htb]
\label{Fig.1}
$$\hspace{0.2cm} \mbox{
\epsfxsize 14.0 true cm
\epsfysize 18.0 true cm
\setbox0= \vbox{
\hbox {
\epsfbox{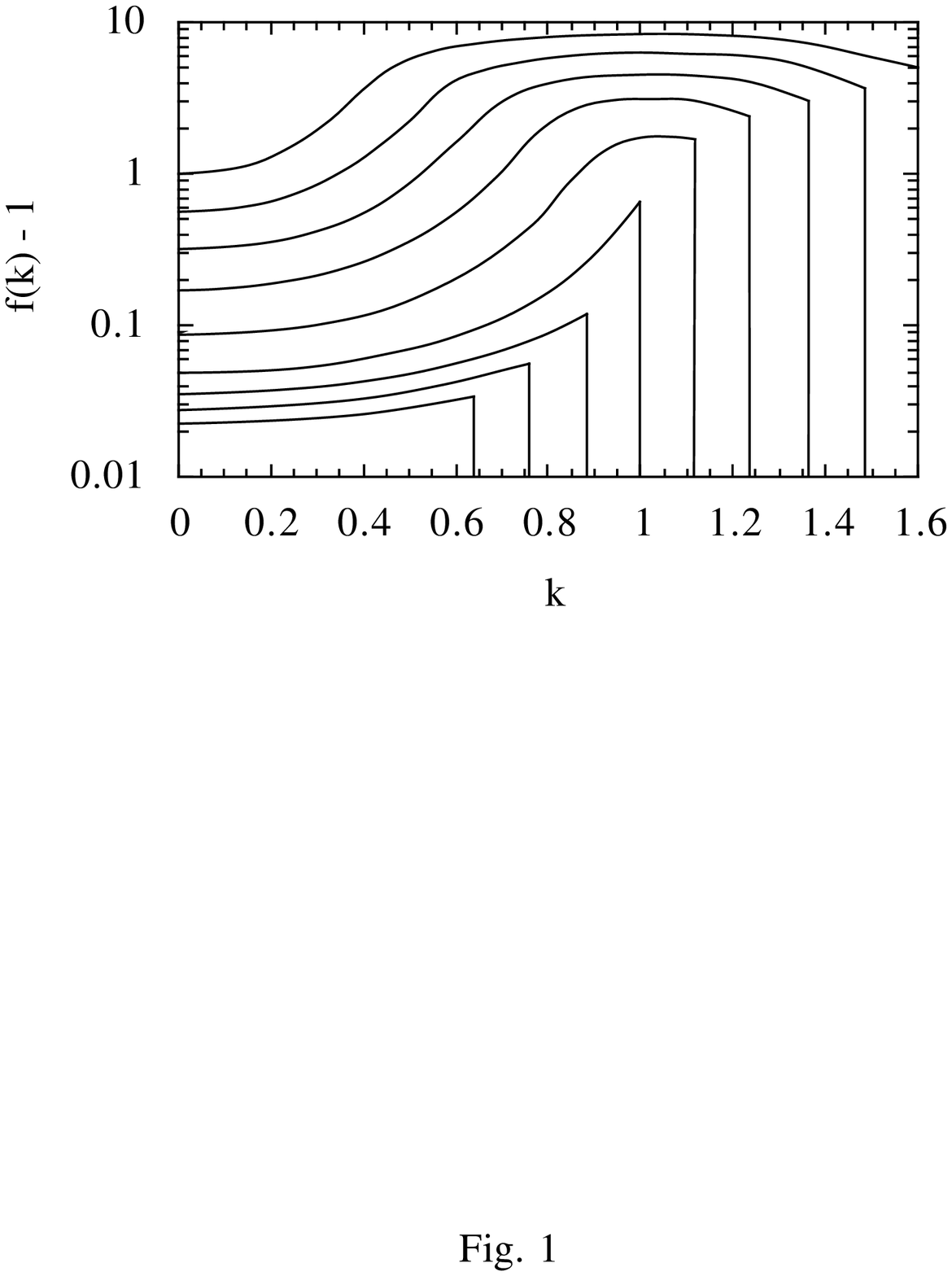}
} 
} 
\box0
} $$
\end{figure}

\begin{figure}[htb]
\label{Fig.2}
$$\hspace{0.2cm} \mbox{
\epsfxsize 15.0 true cm
\epsfysize 18.0 true cm
\setbox0= \vbox{
\hbox {
\epsfbox{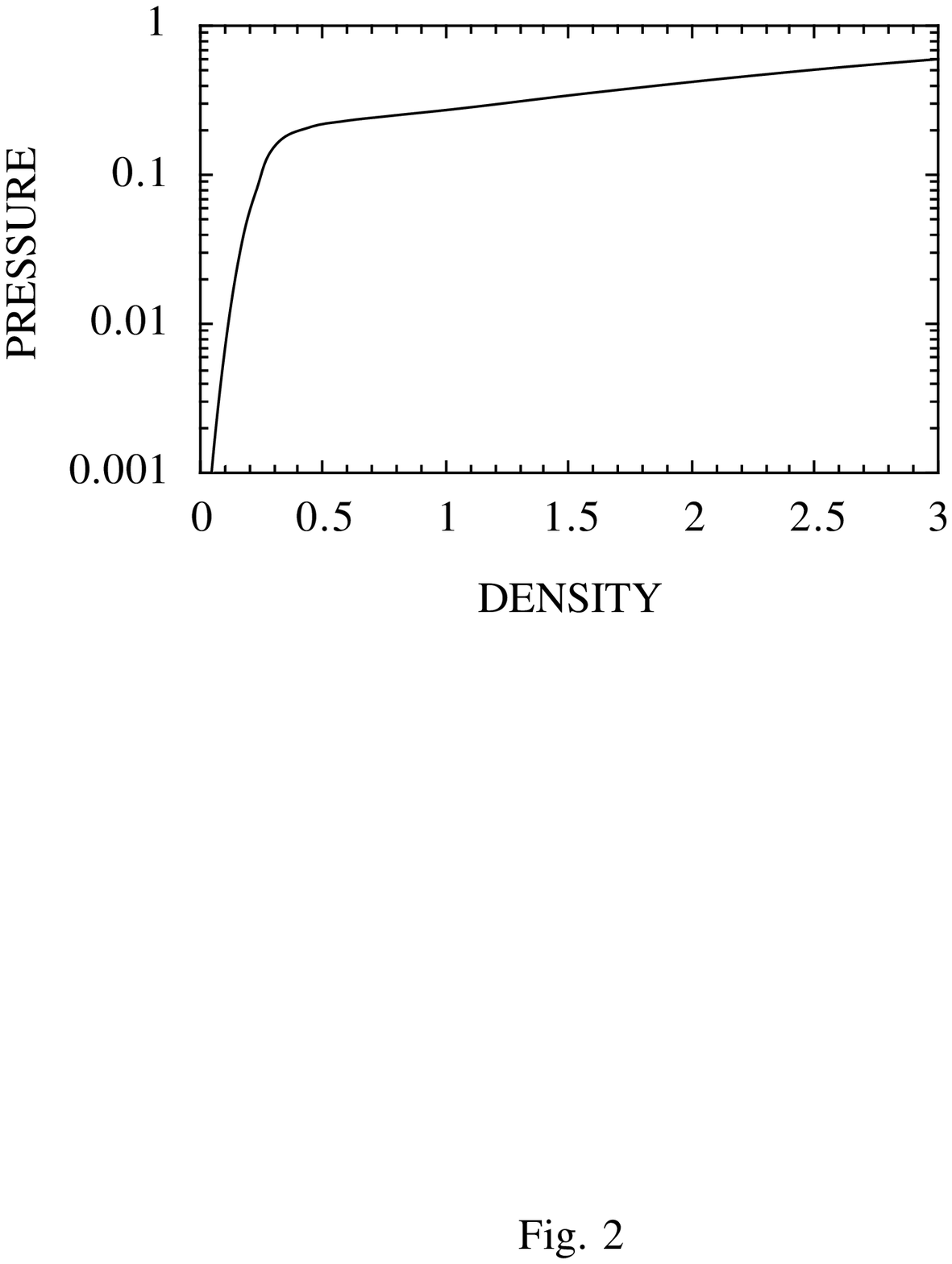}
} 
} 
\box0
} $$
\end{figure}

\end{document}